%% file: main.tex
\documentclass[%
 reprint,
 amsmath,amssymb,
 aps,
]{revtex4-2}

\usepackage{graphicx}
\usepackage{dcolumn}
\usepackage{bm}
\usepackage[utf8]{inputenc}
\usepackage{physics}
\usepackage{quantikz}
\usetikzlibrary{shapes, positioning,calc,fit,decorations.pathreplacing}
\usepackage{tikz}       
\usepackage{amsmath, braket}
\usetikzlibrary{decorations.pathreplacing,calc} 
\usepackage[colorlinks=true,
            linkcolor=blue,
            citecolor=blue,
            urlcolor=blue]{hyperref}
\usepackage[sort&compress]{natbib}

\begin{document}

\preprint{APS/123-QED}

\title{Quantum walk based Monte Carlo simulation\\for photon interaction cross sections}

\author{Euimin Lee\textsuperscript{1,2}}
 \altaffiliation{euimin97@yonsei.ac.kr}
\author{Sangmin Lee\textsuperscript{3}}
 \altaffiliation{sangminLEE@yuhs.ac}
\author{Shiho Kim\textsuperscript{1,2}}%
 \altaffiliation{shiho@yonsei.ac.kr}
\affiliation{\textsuperscript{1}School of Integrated Technology, Yonsei University, Incheon 21983, Republic of Korea \\
\textsuperscript{2}BK21 Graduate Program in Intelligent Semiconductor Technology, Yonsei University, Incheon 21983, Republic of Korea \\
\textsuperscript{3}Department of Radiation Oncology, Yonsei Cancer Center, Yonsei University College of Medicine, Seoul 03722, Republic of Korea
}

\date{\today}

\begin{abstract}
  High-energy physics simulations traditionally rely on classical Monte Carlo methods to model complex particle interactions, often incurring significant computational costs. In this paper, we introduce a novel quantum-enhanced simulation framework that integrates discrete-time quantum walks with quantum amplitude estimation to model photon interaction cross sections. By mapping the probabilistic transport process of 10 MeV photons in a water medium onto a quantum circuit and focusing on Compton scattering as the dominant attenuation mechanism, we demonstrate that our approach reproduces classical probability distributions with high fidelity. Simulation results obtained via the IBM Qiskit quantum simulator reveal a quadratic speedup in amplitude estimation compared to conventional Monte Carlo methods. Our framework not only validates the feasibility of employing quantum algorithms for high-energy physics simulations but also offers a scalable pathway toward incorporating multiple interaction channels and secondary particle production. These findings underscore the potential of quantum-enhanced methods to overcome the computational bottlenecks inherent in large-scale particle physics simulations.
\end{abstract}

\maketitle


\section{\label{sec:level1}Introduction}

High-energy physics (HEP) seeks to uncover the fundamental constituents of matter and their interactions.
However, simulating the intricate processes underlying particle interactions remains a formidable challenge.
Classical Monte Carlo (MC) methods, which underpin many simulation tools like Geant4, rely on generating
millions to billions of random events to accurately predict particle spectra, interaction cross sections, and energy
budgets. This enormous computational burden not only slows down research progress, but also restricts the
exploration of complex experimental conditions, posing a serious constraint on advancing HEP studies.

To alleviate these computational bottlenecks, numerous approaches have been proposed, including parallel computing \cite{1_1, 6_1, 7_1}, machine learning \cite{2_1, 4_1, 10_1, 11_1, 3_1, 8_1}, and advanced sampling techniques \cite{5_1, 9_1}. While these methods offer improvements, they are fundamentally limited by the intrinsic challenges of classical simulations. Despite substantial research efforts, classical methods still have not completely solved the problem of computational bottlenecks.

In this context, quantum computing has attracted significant attention because of its potential to solve classically challenging problems. By leveraging quantum entanglement and quantum superposition, quantum computing offers a fundamentally different approach to simulation by enabling a much more efficient exploration of vast state spaces. In fact, research on quantum algorithms has advanced in various areas of high-energy and nuclear physics such as medium effects in dense matter \cite{12_1, 20_1}, parton showers \cite{13_1, 14_1, 18_1, 25_1}, scattering processes \cite{15_1, 21_1, 63_1}, quantum field simulations \cite{23_1}, event analysis \cite{19_1, 24_1, 26_1}, jet evolution \cite{22_1} and dark sector studies \cite{27_1}. These investigations suggest that the synergy between quantum computing and high-energy physics may lead to innovative computational methods \cite{16_1, 17_1}. Moreover, the versatility of these quantum techniques indicates potential applications in other fields, including finance \cite{32_1, 33_1, 34_1, 35_1, 36_1, 37_1, 62_1}, where complex probabilistic transport models are also of great interest.

Among several quantum algorithms, Quantum Amplitude Estimation (QAE) \cite{29_1} holds the potential to deliver a quadratic speedup over traditional Monte Carlo methods. To apply this approach in practice, classical probability distributions must be mapped to quantum states. This mapping is a significant challenge that impedes the practical implementation of the technique.

In this paper, we propose an integrated framework that combines a discrete time quantum walk \cite{30_1, 31_1} with the quantum amplitude estimation algorithm to simulate high-energy photon interactions in water. By embedding the probabilistic structure of particle transport directly into quantum superposition states via the quantum walk, this approach provides a more physically meaningful representation compared to previous overly simplified models \cite{50_5, 51_5} or those based on quantum generative adversarial networks \cite{52_5, 53_5, 54_5} and reduces the number of samples required to achieve a given level of accuracy.

In our study, we implemented a simplified model for the interaction of 10 MeV photons in water using the IBM Qiskit quantum simulator. The results indicate that our simulation produces very low mean squared error and Kullback-Leiber divergence values compared with classical Monte Carlo methods. This confirms that the combination of the quantum walk and iterative quantum amplitude estimation provides high accuracy and stability in practice. In particular, our implementation of iterative quantum amplitude estimation reproduces the theoretically guaranteed quadratic speedup and exhibits rapid convergence in experiments. In addition the extension of the quantum walk based interaction model demonstrates consistency with classical simulations for large high-energy physics problems and thus proves the method has robust scalability. Our study can be regarded as a proof of concept for simulating general particle transport properties beyond photon cross section estimation.

The current model assumes a single interaction channel with complete photon absorption, reflecting current hardware limitations. In the future, it will be necessary to incorporate multiple interaction channels, partial energy loss and secondary particle generation to fully capture the complexity of realistic high-energy processes.

This paper is organized as follows. In Sec.\,\hyperref[sec:Preliminary]{II}, we briefly review the theoretical background of quantum walks (QW) and quantum amplitude estimation (QAE). In Sec.,\hyperref[sec:QW_algorithm]{III}, we describe the proposed quantum simulation algorithm based on quantum walk and quantum amplitude estimation for photon interactions. In Sec.\,\hyperref[sec:QAE_integration]{IV}, we present simulation results using Qiskit and demonstrates the speedup achieved relative to classical methods. In Sec.\,\hyperref[sec:discussion]{V} we analyze the results and discuss  comparisons with existing approaches, limitations and potential directions for future research. Finally, Sec.\,\hyperref[sec:conclusion]{VI} summarizes the overall achievement of the paper.

\input{chapter2}

\input{chapter3}

\input{chapter4}

\input{chapter5}

\begin{acknowledgments}
This research was supported by \textit{Quantum Computing based on Quantum Advantage challenge research (RS-2023-00257561)} through the National Research Foundation of Korea (NRF) funded by the Korean government (Ministry of Science and ICT (MSIT)).

\end{acknowledgments}

\section*{Data Availability}
The data that support the findings of this article are not publicly available upon publication because the cost of preparing, depositing, and hosting the data would be prohibitive within the terms of this research project. The data are available from the authors upon reasonable request.

\bibliography{main}
\end{document}

%% file: chapter2.tex
\section{Theoritical Background}
\label{sec:Preliminary}
\subsection{Quantum Walk}

Quantum random walk (QW) \cite{30_1,31_1} is the quantum analog of classical random walk. 
It describes the motion of a particle known as a walker that occupies specific positions on a graph. 
In a discrete-time QW, each step involves a coin flip that determines the walker's direction of travel. 
The walker's state is specified by the position $x$ and the coin outcome $c$, written as $\ket{x,c}$. 
By repeatedly applying two operations, the coin operation $C$ that selects the direction and the shift operation $S$ 
that moves the walker to the next position, one can simulate a probabilistic process in a fully quantum-mechanical framework.

Compared to the classical random walk, whose shift operation places the walker at one fixed position based on the coin result, 
the quantum random walk employs a quantum coin such as the Hadamard coin to create a superposition of basis states in the coin space $H_C$. 
When measured, the wave function collapses to a single outcome, mirroring the classical random walk. 
For example, the Hadamard coin can be written as
\begin{equation}
H \;=\;
\frac{1}{\sqrt{2}}
\begin{pmatrix}
1 & 1 \\
1 & -1
\end{pmatrix},
\end{equation}
and measuring this coin yields two basis states with equal probability.

In our proposed QW-based algorithm, we use a coin formed by a rotation around the $y$ axis, often denoted $R_y(\theta)$. 
This $R_y$ gate acts as the coin operation. Combined with the shift operation, it forms a single unitary transformation 
that acts on the qubit state,
\begin{equation}
U = S \bigl(C \otimes I\bigr.
\bigl).
\end{equation}
Applying $U$ iteratively over multiple steps gives a series of position updates governed by quantum superposition 
and interference effects.

The coin and shift structure of the quantum random walk is well suited to simulating interaction models 
that involve stochastic processes. In particle physics, Markov chain approaches are frequently used to handle 
branching probabilities for emission, scattering, or absorption, as seen in parton showers or radiation transport. 
A QW encodes these interaction probabilities in the coin operation and the resulting state transitions in the shift operation, 
capturing probabilistic branching in a quantum framework. In the experiments presented in this paper, we use a two-dimensional coin space $\mathcal{H}_C$ to minimize the quantum circuit volume for efficient implementation. However, one can expand the dimension of $\mathcal{H}_C$ by $\log_{2}(k)$, thus enabling the simulation of $k$ events at each step. 
As a result, the state
\begin{equation}
\Bigl[S \bigl(C \otimes I\bigr)\Bigr]^N \ket{x_0, c_0}
\end{equation}
can simulate a discrete stochastic process of size $k^N$ in a purely quantum system \cite{32_1}.

In summary, a quantum random walk combines the coin operation for probabilistic branching 
and the shift operation for position updates in a unitary way. 
Compared to classical random walk, quantum walk displays richer interference effects and faster spreading. 
As a result, they have attracted attention as a powerful tool for modeling complex probabilistic processes, 
for example radiation transport, with broad potential for diverse applications.

\subsection{Quantum Amplitude Estimation}

Quantum Amplitude Estimation (QAE) \cite{29_1} is an extension of Grover's algorithm \cite{38_2,39_2} for amplitude estimation tasks, which provides a quadratic speed up compared to the traditional Monte Carlo method on classical computers. In this algorithm, the problem of interest is given by a unitary operator $A$ acting on $n+1$ qubits, assuming the following condition:
\begin{equation}
A \lvert 0\rangle^n \lvert 0\rangle
= \sqrt{1-a}\,\lvert \Psi_0\rangle^n \lvert 0\rangle
+ \sqrt{a}\,\lvert \Psi_1\rangle^n \lvert 1\rangle,
\end{equation}

\input{fig1.tex}

\setlength\abovedisplayskip{10pt plus 2pt minus 2pt}
\setlength\belowdisplayskip{10pt plus 2pt minus 2pt}

\noindent where $a \in [0,1]$, and $\lvert \Psi_0\rangle$ and $\lvert \Psi_1\rangle$ are two orthonormal states.

To estimate $a$, we define the Grover operator $Q$ as
\begin{equation}
Q = A\, S_0\, A^\dagger \,S_{\psi_1},
\end{equation}
where
$S_0 = I - 2\,\lvert 0\rangle^{n+1}\langle 0\rvert^{n+1}$ 
and 
$S_{\psi_1} = I - 2\,\lvert \psi_1\rangle\langle \psi_1\rvert 
            \otimes \lvert 1\rangle\langle 1\rvert$
are sign flipping operators.

The standard form of QAE \cite{29_1} is derived from Quantum Phase Estimation (QPE). In Fig.~\ref{fig:QPE}(a), QPE circuit with $m$ ancillary qubits applies $Q^k$ in an exponentially increasing powers to estimate the amplitude $a$ as 
\begin{equation}
\bar{a} = \frac{y\,\pi}{2^m} 
\quad \text{for} \quad
y \in \{0,\dots, 2^m - 1\}.
\end{equation}

In this case, the estimation error 
\(\epsilon = \lvert a - \bar{a}\rvert\)
satisfies
\begin{equation}
\lvert a - \tilde{a} \rvert
\;\le\;
\frac{2\sqrt{\,a(1-a)\,}\,\pi}{M} 
+\;
\frac{\pi^2}{M^2}
\;=\;
O(M^{-1}),
\end{equation}
indicating a quadratic speedup compared to \(O(M^{-1/2})\) in classical Monte Carlo methods.

The standard QAE approach uses ancillary qubits and the QFT, and consequently the resulting estimate \(\tilde{a}\) is restricted to a discrete grid. To overcome this, various modified methods have recently been proposed \cite{40_2, 41_2, 42_2, 43_2} to reduce circuit complexity and obtain estimates over a continuous range. These methods remove the need for ancillary qubits and QFT by directly applying the operation \(Q^k A\) for amplitude estimation. Let \(a = \sin^2(\theta_a)\). Then
\begin{equation}
\begin{aligned}
Q^k\, A \,\lvert 0\rangle_n \lvert 0\rangle
&= \cos\bigl((2k + 1)\,\theta_a\bigr)\,\lvert \psi_0\rangle_n \lvert 0\rangle \\
&\quad + \sin\bigl((2k + 1)\,\theta_a\bigr)\,\lvert \psi_1\rangle_n \lvert 1\rangle,
\end{aligned}
\end{equation}
 where the probability of measuring \(\lvert 1\rangle\) is \(\sin^2\bigl((2k+1)\,\theta_a\bigr)\). By selecting different values of \(k\) and combining their outcomes, one can achieve an error bound similar to that of QPE. Each algorithm differs in how it chooses the exponent \(k\) of the Grover operator \(Q\) and how it aggregates the measurement outcomes into the final estimate of \(a\). In this work, we adopt the Iterative Quantum Amplitude Estimation (IQAE) \cite{40_2} method . Fig.~\ref{fig:QPE}(b) shows a generic circuit example for these QPE-free QAE approaches.

%% file: fig1.tex
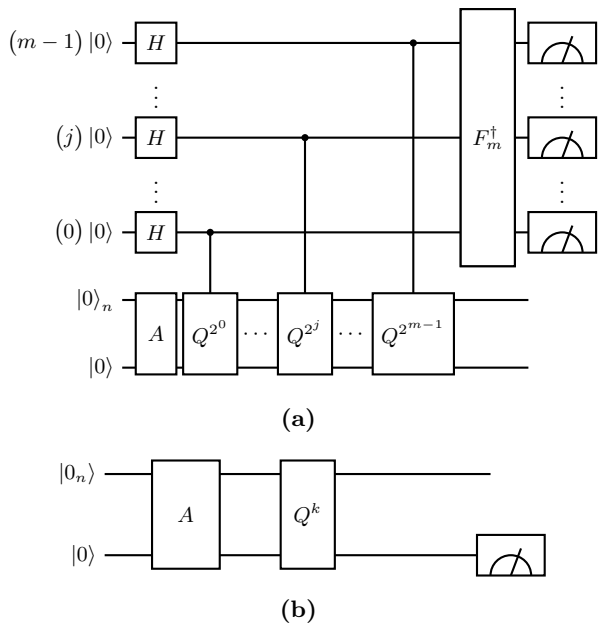
\begin{figure}[ht]
    \centering
    
    \begin{minipage}[b]{0.45\textwidth}
        \centering
       \begin{tikzpicture}[thick, scale=0.9, every node/.style={transform shape}]

\tikzset{
  measure/.style={
    rectangle,
    draw,
    minimum width=1cm,
    minimum height=0.6cm,
    path picture={
      \draw (path picture bounding box.south west) ++(0.2,0)
            arc [start angle=180, end angle=0, radius=0.3];
      \draw (path picture bounding box.south west) ++(0.5,0)
            -- ++(0.15,0.4);
    }
  }
}


\newcommand{\yTop}{0}        
\newcommand{\yDotsTop}{-0.7} 
\newcommand{\yMid}{-1.4}     
\newcommand{\yDotsBot}{-2.1} 
\newcommand{\yBot}{-2.8}     
\newcommand{\yRegN}{-3.8}    
\newcommand{\yRegOne}{-4.8}

\newcommand{\xStart}{-0.5}
\newcommand{\xEnd}{5.5}
\newcommand{\xH}{0}        
\newcommand{\xFblock}{5.5} 
\newcommand{\xFblockEnd}{5}
\newcommand{\xMeasure}{6} 

\newcommand{\xA}{0}      
\newcommand{\xQone}{0.8}   
\newcommand{\xQtwo}{2.2}   
\newcommand{\xQthree}{3.8} 


\draw (\xStart,\yTop) -- (\xEnd,\yTop);
\node[left] at (\xStart,\yTop) {$\bigl(m-1\bigr)\ket{0}$};

\node at (0,\yDotsTop) {$\vdots$};
\node at (\xMeasure, \yDotsTop) {$\vdots$};

\draw (\xStart,\yMid) -- (\xEnd,\yMid);
\node[left] at (\xStart,\yMid) {$\bigl(j\bigr)\ket{0}$};

\node at (0,\yDotsBot) {$\vdots$};
\node at (\xMeasure, \yDotsBot) {$\vdots$};

\draw (\xStart,\yBot) -- (\xEnd,\yBot);
\node[left] at (\xStart,\yBot) {$\bigl(0\bigr)\ket{0}$};

\draw (\xStart,\yRegN) -- (\xEnd,\yRegN);
\node[left] at (\xStart,\yRegN) {$\ket{0}_n$};

\draw (\xStart,\yRegOne) -- (\xEnd,\yRegOne);
\node[left] at (\xStart,\yRegOne) {$\ket{0}$};


\draw[fill=white] (\xH-0.3,\yTop+0.3) rectangle (\xH+0.3,\yTop-0.3);
\node at (\xH,\yTop) {$H$};

\draw[fill=white] (\xH-0.3,\yMid+0.3) rectangle (\xH+0.3,\yMid-0.3);
\node at (\xH,\yMid) {$H$};

\draw[fill=white] (\xH-0.3,\yBot+0.3) rectangle (\xH+0.3,\yBot-0.3);
\node at (\xH,\yBot) {$H$};


\draw[fill=white] (\xA-0.3,\yRegN+0.1) rectangle (\xA+0.3,\yRegN-1.1);
\node at (\xA,\yRegN-0.5) {$A$};

\draw[fill=white] (\xQone-0.4,\yRegN+0.1) rectangle (\xQone+0.4,\yRegN-1.1);
\node at (\xQone,\yRegN-0.5) {$Q^{2^0}$};
\fill (\xQone,\yBot) circle (1.5pt);
\draw (\xQone,\yBot) -- (\xQone,\yRegN+0.1);

\draw[fill=white] (\xQtwo-0.4,\yRegN+0.1) rectangle (\xQtwo+0.4,\yRegN-1.1);
\node at (\xQtwo,\yRegN-0.5) {$Q^{2^j}$};
\fill (\xQtwo,\yMid) circle (1.5pt);
\draw (\xQtwo,\yMid) -- (\xQtwo,\yRegN+0.1);

\draw[fill=white] (\xQthree-0.6,\yRegN+0.1) rectangle (\xQthree+0.6,\yRegN-1.1);
\node at (\xQthree,\yRegN-0.5) {$Q^{2^{m-1}}$};
\fill (\xQthree,\yTop) circle (1.5pt);
\draw (\xQthree,\yTop) -- (\xQthree,\yRegN+0.1);


\node at ( \xQone/2 + \xQtwo/2, \yRegOne +0.5 ) {$\cdots$};

\node at ( \xQthree/2  + \xQtwo/2 - 0.1 , \yRegOne + 0.5 ) {$\cdots$};


\draw[draw, fill=white]  (4.5, \yTop+0.5) rectangle (5.3, \yBot-0.5);
\node at ( \xQthree  + 1.1 , \yMid ) {$F_m^\dagger$};

\foreach \ycoord in {\yTop, \yMid, \yBot}{
   \draw (4.5,\ycoord) -- (5.3,\ycoord);
}


\foreach \ycoord in {\yTop, \yMid, \yBot}{
  \node[measure] at (\xMeasure,\ycoord) {};
}

\draw[draw, fill=white]  (4.5, \yTop+0.5) rectangle (5.3, \yBot-0.5);
\node at ( \xQthree  + 1.1 , \yMid ) {$F_m^\dagger$};

\end{tikzpicture}
        
        \vskip 0.5em
        \centering
        \textbf{(a)} 
    \end{minipage}
\hspace{0.1\textwidth}
    
    \begin{minipage}[b]{0.45\textwidth}
        \centering
        \begin{tikzpicture}[thick, scale=0.9, every node/.style={transform shape}]

\tikzset{
  measure/.style={
    rectangle,
    draw,
    fill=white,      
    minimum width=1cm,
    minimum height=0.6cm,
    path picture={
      \draw (path picture bounding box.south west) ++(0.2,0)
            arc [start angle=180, end angle=0, radius=0.3];
      \draw (path picture bounding box.south west) ++(0.5,0)
            -- ++(0.15,0.4);
    }
  }
}


\newcommand{\ySN}{0}      
\newcommand{\yAnc}{-1.2}  
\newcommand{\xStart}{0}
\newcommand{\xEnd}{5.7}

\draw (\xStart,\ySN) -- (\xEnd,\ySN);
\node[left] at (\xStart,\ySN) {$\lvert 0_n \rangle$};

\draw (\xStart,\yAnc) -- (\xEnd,\yAnc);
\node[left] at (\xStart,\yAnc) {$\lvert 0\rangle$};


\newcommand{\xA}{1.2}
\newcommand{\xQ}{3.0}
\newcommand{\xMeasure}{6}

\draw[draw, fill=white] (\xA-0.5,\ySN+0.2) rectangle (\xA+0.5,\ySN-1.4);
\node at (\xA,\ySN-0.6) {$A$};

\draw[draw, fill=white] (\xQ-0.4,\ySN+0.2) rectangle (\xQ+0.4,\ySN-1.4);
\node at (\xQ,\ySN-0.6) {$Q^k$};


\node[measure] at (\xMeasure,\yAnc) {};


\end{tikzpicture}
        
        \vskip 0.5em
        \centering
        \textbf{(b)} 
    \end{minipage}

    \caption{Quantum amplitude estimation using (a) QPE circuit and (b) QPE-free circuit.}
    \label{fig:QPE}
\end{figure}

%% file: chapter3.tex
  \section{Quantum Algorithm for Photon Interaction Simulation}
\label{sec:QW_algorithm}

\subsection{Theoretical outline of the particle interaction algorithm}

 In high-energy physics (HEP) simulations of particle trajectories, each interaction step is determined by a probabilistic branching process. This approach is essentially a Markov chain, in which possible events occur with specific probabilities depending on the current state. For instance, classical Monte Carlo simulations track the final trajectory by randomly choosing from these probabilistic branches at each step, repeating many trials to accumulate sufficient statistics.

The exponential attenuation law describing the photon intensity $I(x)$ transmitted through a material of thickness $x$ is given by
\begin{equation}
I(x) = I_0 \, e^{-\mu x},
\end{equation}
where $I_0$ is the initial photon intensity and $\mu$ is the linear attenuation coefficient. Dividing the distance into segments of length $\Delta x$, the probability $p$ of having at least one interaction in that segment is
\begin{equation}
p = 1 - e^{-\mu \,\Delta x}.
\end{equation}

When a photon with energies ranging from a few MeV up to several tens of MeV traverses the material, multiple interaction mechanisms may occur. Examples include:
\begin{itemize}
    \item Photoelectric effect, $\sigma_{\mathrm{pe}}$,
    \item Coherent (Rayleigh) scattering, $\sigma_{\mathrm{coh}}$,
    \item Incoherent (Compton) scattering, $\sigma_{\mathrm{incoh}}$,
    \item Pair production, $\sigma_{\mathrm{pair}}$,
    \item Triplet production, $\sigma_{\mathrm{trip}}$,
    \item Photonuclear reactions, $\sigma_{\mathrm{ph.n.}}$.
\end{itemize}

Each channel contributes in the form of a cross section $\sigma_i(E_\gamma)$; summing them yields
\begin{equation}
\mu(E_\gamma)
= \rho \,\Bigl[\sigma_{\mathrm{pe}}(E_\gamma)
+ \sigma_{\mathrm{coh}}(E_\gamma)
+ \sigma_{\mathrm{incoh}}(E_\gamma) + \dots \Bigr],
\end{equation}
where $\rho$ is the material density. For a 10\,MeV photon incident on water, contributions other than Compton scattering and pair production amount to less than 1\%, making them negligible \cite{44_3}. Since Compton scattering alone accounts for over 75\% of the total attenuation, the subsequent calculations focus exclusively on Compton scattering. This simplification was essential for designing circuits that can run on quantum simulators with qubit limitations, representing a deliberate trade-off between physical realism and the constraints of quantum simulator. Hence,
\begin{equation}
\mu(E_\gamma)
\,\approx\,
\Sigma_{\mathrm{Compton}}(E_\gamma)
=
\rho \,\sigma_{\mathrm{Compton}}(E_\gamma).
\end{equation}

The total Compton cross section $\sigma_{\mathrm{Compton}}(E_\gamma)$ can be obtained by integrating the Klein--Nishina differential cross section \cite{45_3}. Specifically, if $E_\gamma$ is the photon energy and $\theta$ the scattering angle,
\begin{equation}
\frac{d\sigma}{d\Omega}(E_\gamma,\theta)
= r_e^2 \,\Bigl(\tfrac{E_\gamma'}{E_\gamma}\Bigr)^2
\Bigl(\tfrac{E_\gamma'}{E_\gamma} + \tfrac{E_\gamma}{E_\gamma'} - \sin^2\theta\Bigr),
\end{equation}
\begin{equation}
E_\gamma' =
\frac{E_\gamma}{\,1 + \tfrac{E_\gamma}{m_e c^2}\,(1-\cos\theta)\,},
\end{equation}
where $r_e$ is the classical electron radius and $E_\gamma'$ is the scattered photon energy. Integrating over $\theta \in [0,\pi]$ yields the total Compton cross section:
\begin{equation}
\sigma_{\mathrm{Compton}}(E_\gamma)
= \int_{0}^{\pi}
\left(\frac{d\sigma}{d\Omega}\right)
2\pi \,\sin\theta \, d\theta.
\end{equation}

\begin{figure}{t}
    \centering
    \includegraphics[width=0.5\textwidth]{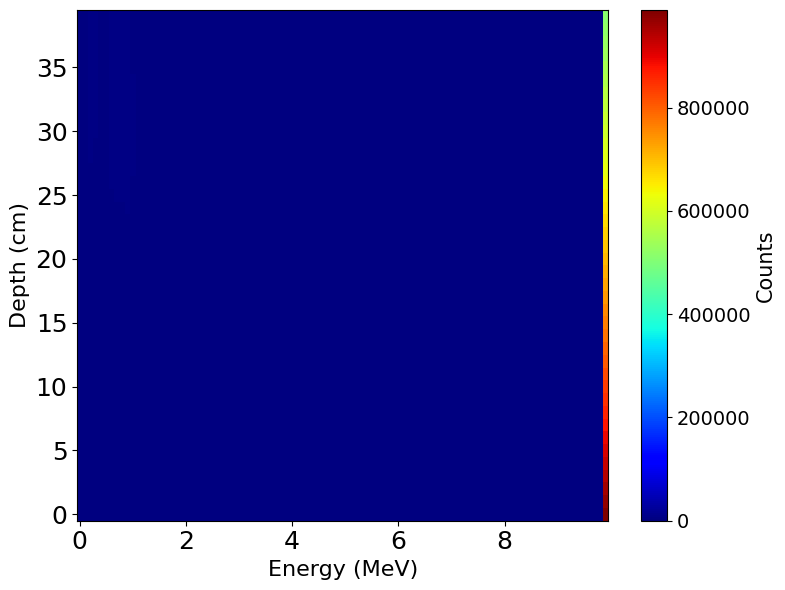}
    \caption{%
    Depth-wise energy spectrum histogram for a 10\,MeV photon beam incident on a water medium, 
    taking into account only Compton scattering. 
    The simulation was performed with a depth of 1\,cm 
    and energy bins of 0.1\,MeV.
    }
    \label{fig:depth_spectrum}
\end{figure}

In modeling the interaction of high-energy photons in matter, Compton scattering typically leads to a continuous distribution of residual energies, depending on the scattering angle. Meanwhile, if one considers only Compton scattering for a 10\,MeV photon passing through a water medium, partial energy-loss events occur with extremely low probability, rendering their impact on the overall attenuation profile effectively negligible. According to classical Monte Carlo simulations performed with GEANT4 \cite{46_3,47_3}, most events retain approximately 10\,MeV, while only a small fraction undergoes some degree of energy loss. This is illustrated in Fig.~\ref{fig:depth_spectrum} by an energy spectrum histogram.

Based on these observations, we adopt a simplified one-dimensional binary model in which the photon remains at 10 MeV if no interaction occurs, and is treated as fully absorbed otherwise. By omitting the numerous
low-probability channels associated with partial energy downgrades,
we substantially reduce the complexity of the quantum circuit and implementation becomes possible \cite{48_3}. Although this approach does not
reproduce the entire spectrum, it captures the dominant
interaction features at 10\,MeV. Our preliminary analysis indicates
that the probability mass of intermediate-energy states is negligible
for the purpose of this study.

This can decrease the circuit width by a factor of $\log_{2}(n)$
compared to a model with $n$ separate probabilistic branches.Since fully implementing every branch on IBM’s 32-qubit quantum simulator is not yet feasible due to the limited qubit count, we employ this simplified toy model. This approach is sufficient for proof-of-concept purposes, and once hardware capabilities improve, one can extend the number of possible interaction outcomes per branch from two to an arbitrary integer 
\(k\), enabling more general simulations.

\subsection{Quantum walk implementation of photon interaction}

The photon interaction model is implemented via the quantum random walk 
algorithm as outlined in Sec.\,\textsc{II}. In this approach, the coin 
operation is defined as a unitary rotation corresponding to the interaction 
probability \(p_k\) computed at the \(k\)-th step. The operator takes the form
\begin{equation}
U_{k} \;=\;
\begin{pmatrix}
\sqrt{\,1 - p_k}\, & \;-\sqrt{\,p_k}\,\\[6pt]
\sqrt{\,p_k}\,     & \;\sqrt{\,1 - p_k}\,
\end{pmatrix},
\end{equation}
where $p_{k}$ is the probability that the photon at step $k$ interacts and loses all its energy. 
The coin space \(\mathcal{H}_C\) is two-dimensional, consisting of the two quantum states $\{|0\rangle, |1\rangle\}$. 
In this experiment, we define $|0\rangle$ as the ``no interaction'' state and $|1\rangle$ as the ``interaction occurs'' state.

\begin{figure}[t]
\centering
\begin{tikzpicture}[scale=1.0, every node/.style={font=\small}]

\draw (0,0) -- (7,0);
\draw (0,-1) -- (7,-1);
\node at (-0.4,-2) {$|c\rangle$};
\draw (0,-2) -- (7,-2);
\node at (-0.4,-3) {$|w\rangle$};
\draw (0,-3) -- (7,-3);

\draw[decorate, decoration={brace,mirror,amplitude=5pt}]
  (-0.2,0.2) -- (-0.2,-1.2) node[midway,left=3pt]{$|x\rangle$};

\fill (2,0) circle (2pt);
\fill (2,-1) circle (2pt);
\draw (2,0)--(2,-3);
\draw[fill=white] (1.7,-3.3) rectangle (2.3,-2.7);
\node at (2,-3) {\large W};

\node at (4,-0.4) {\huge\(\vdots\)};
\node at (1,-0.4) {\huge\(\vdots\)};

\fill (3,-3) circle (2pt);
\draw (3,-3)--(3,-2);
\draw[fill=white] (2.7,-2.3) rectangle (3.3,-1.7);
\node at (3,-2) {\large C};

\draw[dashed, thin] (1.4,0.4) rectangle (3.6,-3.4);
\node at (2.2,0.65) {{Position check and operation}};

\fill (5,-2) circle (2pt);
\draw (5,-2)--(5,-1);
\draw[fill=white] (4.7,0.1) rectangle (5.3,-1.3);
\node at (5,-0.6) {\large S};

\draw[dashed,thin] (4.4,0.4) rectangle (5.6,-2.2);
\node at (5,0.65) {{Shift}};

\end{tikzpicture}
\caption{%
Schematic for a single step of a quantum walk algorithm modeling photon interactions in a simplified scenario. The “position check” unit reads the walker’s current depth. The coin operation applies the appropriate interaction probability, while the shift operation advances or halts the walker depending on the coin result.
}
\label{fig:qc-walk}
\end{figure}
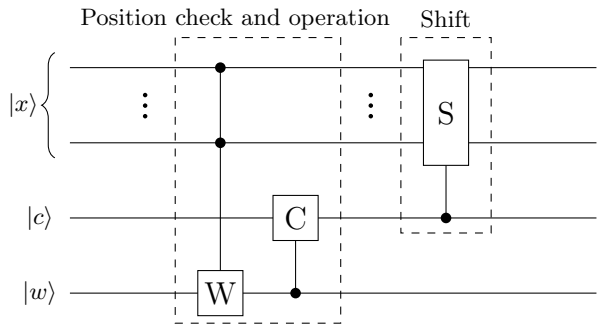

Next, the position space \(\mathcal{H}_P\) discretizes the photon's current location in increments of 1\,cm, allowing only nonnegative integer states 
\(\{\lvert i\rangle \,:\, i \in \mathbb{N}_{0}\}\). The shift operation 
is controlled by the coin qubit and moves the walker in the correct direction. 
Fig.~\ref{fig:qc-walk} illustrates a conceptual diagram of how the quantum 
walk can represent a real photon interaction process. The current photon’s depth 
is stored in the walker’s position, initialized at \(\lvert 0\rangle\). Using 
\(x\) qubits for the position register allows up to \(2^x\) discrete depth 
levels, offering exponential scalability. During each step, a \emph{position check} 
procedure applies the appropriate coin operation with probability \(p_k\) 
based on the position of the walker, typically implemented through multiple
\textsc{CCNOT} gates to ensure that the operation is fully unitary.

Afterwards, the shift operation updates the position of the photon depending
on the coin result. If the coin outcome is \(\lvert 0\rangle\), we assume that the photon advances by 1\,cm, so the walker increments its position by 1. Conversely, if the coin outcome is \(\lvert 1\rangle\), 
the photon loses all its energy at that depth, and the walker remains in place. 
Repeating these steps for the desired number of discrete depth intervals completes 
the quantum-walk-based simulation of the photon’s interaction. 

\subsection{Integrating QAE with Quantum Walk for photon interaction simulation}

Based on the quantum random walk-based photon interaction model described above, we simplify the interaction outcomes in each branching step of photon transport into two cases. The first case is when no interaction occurs, and the second is when an interaction happens, resulting in complete energy loss. However, as in classical Monte Carlo, a single simulation still suffers from statistical uncertainty, so achieving high accuracy requires multiple circuit repetitions and/or measurements. In such cases, Quantum Amplitude Estimation (QAE) can reduce the number of repetitions needed to estimate the target physical quantity. For example, if we define the state $\lvert \psi_1 \rangle$ to represent “the photon has passed beyond a certain depth $x$ and has not yet interacted,” then the amplitude (and its square) of this state corresponds to the probability that the photon survives beyond depth $x$. By employing QAE, one can theoretically realize a quadratic speed up \cite{29_1}, moving from $O(1/\epsilon^2)$ classical sampling to $O(1/\epsilon)$ queries to estimate that probability.

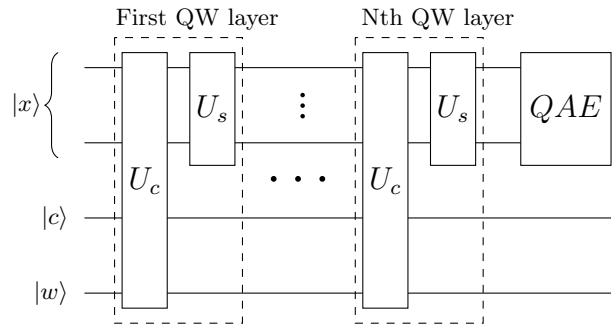
\begin{figure}[t]
\centering
\begin{tikzpicture}[scale=1.0, every node/.style={font=\small}]

\draw (0,0) -- (7,0);
\draw (0,-1) -- (7,-1);

\node[left] at (-0.1,-2) {$|c\rangle$};
\draw (0,-2) -- (7,-2);

\node[left] at (-0.1,-3) {$|w\rangle$};
\draw (0,-3) -- (7,-3);

\draw[decorate, decoration={brace,mirror,amplitude=5pt}]
  (-0.35,0.2) -- (-0.35,-1.2) node[midway,left=3pt]{$|x\rangle$};

\draw[dashed, thin] (0.4,0.4) rectangle (2.1,-3.4);
\node at (1.5,0.65) {{First QW layer}};


\draw[fill=white] (0.5 ,0.2) rectangle (1.1,-3.2);
\node at (0.8,-1.5) {\large $U_c$};

\node at (2.9,-1.5) {\huge $\cdots$};

\draw[fill=white] (1.4,0.2) rectangle (2.0,-1.3);
\node at (1.7,-0.55) {\large $U_s$};

\node at (2.9,-0.4) {\huge $\vdots$};

\draw[dashed, thin] (3.6,0.4) rectangle (5.3,-3.4);
\node at (4.7,0.65) {{Nth QW layer}};


\draw[fill=white] (3.7 ,0.2) rectangle (4.3,-3.2);
\node at (4.0,-1.5) {\large $U_c$};

\draw[fill=white] (4.6,0.2) rectangle (5.2,-1.3);
\node at (4.9,-0.55) {\large $U_s$};

\draw[fill=white] (5.8,0.2) rectangle (7.0,-1.3);
\node at (6.4,-0.55) {\large $QAE$};

\end{tikzpicture}
\caption{%
Schematic of our combined framework, where quantum walks (QW) provide 
a stepwise embedding of the photon’s probabilistic evolution into 
a quantum circuit, and Quantum Amplitude Estimation (QAE) further 
enhances the efficiency of extracting key probabilities. 
Each dashed box represents a single QW iteration, consisting of the coin operation 
($U_c$) and shift operation ($U_s$). The final stage is a QAE module 
for amplitude estimation.%
}
\label{fig:qc-walk-extended}
\end{figure}
\noindent

However, there are two important issues to consider before applying QAE. First, in order to input the probability distribution into the quantum circuit for estimation, the Grover operator $Q$ must be a unitary operation. Consequently, one cannot use non-unitary operations like a reset gate. This limitation makes it difficult to reuse registers without expanding the circuit or adding additional control logic. If reset gates were allowed, we would only need a single qubit for the coin space \(\mathcal{H}_C\) in a quantum walk with a two-dimensional coin space, as considered in our simplified model. However, without relying on reset gates, the coin register requires $\log_2 N$ qubits if the dimension of the position space \(\mathcal{H}_P\) is $N$.

Second, real hardware imposes strict limits on the number of available qubits and the allowable circuit depth. In particular, QAE-based on quantum phase estimation (QPE) requires a large number of ancillary qubits. Furthermore, when running on an actual quantum device, deeper circuits accumulate more noise, reducing the practical advantages.

In this work, we address these challenges by mapping a typical discrete stochastic process to a quantum state via a quantum walk. For instance, we can encode whether the photon is absorbed at step $x$ or passes beyond depth $x$ in two registers (coin and position). While a classical Monte Carlo simulation would sample each step, here we replace that procedure with quantum gates. As a result, at the end of the simulation, the system is in a superposition of all possible trajectories. We aim to apply the quantum amplitude estimation algorithm to this quantum state to obtain the desired statistical result.

Meanwhile, various algorithms have been proposed that preserve the benefits of amplitude estimation without relying on QPE. By properly constructing the Grover operators $Q_k^{AF}$, one can still achieve the theoretical speed up. Among these methods, we adopt the Iterative Quantum Amplitude Estimation approach, which gradually adjusts the number of Grover operators at each step to ensure convergence. This obviates the need for many ancillary qubits or excessively deep circuits \cite{40_2}, making it particularly advantageous under hardware constraints where accuracy and confidence intervals must be balanced.

%% file: chapter4.tex
\section{Simulation Results}
\label{sec:QAE_integration}

\begin{figure}[t]
    \centering

    \includegraphics[width=0.5\textwidth]{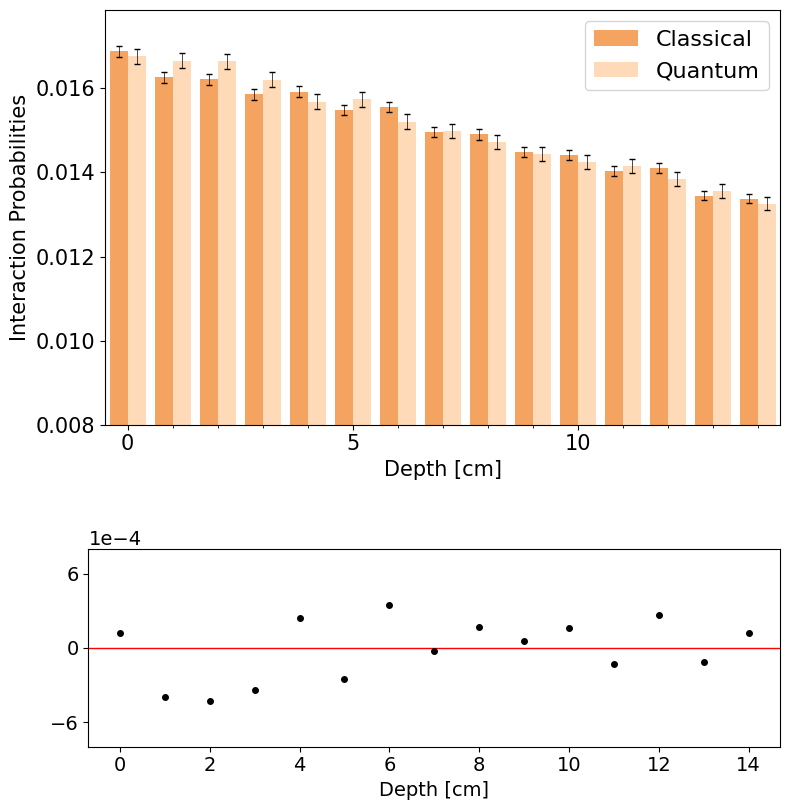}
    \caption{Probability distribution of the 15-step photon interaction simulation for the classical and quantum algorithms. The quantum algorithm has been run on quantum simulator for 500,000 shots, and the classical algorithm
has been run for 1,000,000 shots.}
    \label{fig:two_images}
\end{figure}

In this section, we present the simulation results for the interaction of 10\,MeV photons in a water medium, based on the quantum framework described in Sec.\hyperref[sec:QW_algorithm]{III}. While it is possible to perform such simulations on real quantum computers, the current quantum hardware suffers from limitations such as short coherence time and a restricted number of qubits, making large-scale simulations difficult. Therefore, in this study, we use IBM Qiskit quantum simulator to emulate an ideal quantum device.

Using the physical model described in Sec.\hyperref[sec:QW_algorithm]{III} (a), we compute the photon interaction probability in increments of 1\,cm. In this study, the amplitude of the quantum state is physically associated with the survival probability of a 10 MeV photon after propagating through a specific depth. This directly corresponds to the experimentally measurable transmission rate, enabling the quantum simulation to quantitatively predict the attenuation characteristics of high-energy photons. Since Compton scattering accounts for most of the attenuation of 10\,MeV photons in water, we consider only this scattering mechanism and simplify our model to two outcomes. Accordingly, if no interaction occurs, the photon retains its energy at 10\,MeV; if interaction does occur, the photon is fully absorbed. This simplification is essential to implement the quantum circuit on a 32-qubit simulator given current computational resource constraints.

\begin{figure}[t]
    \centering
    
    \includegraphics[width=0.5\textwidth]{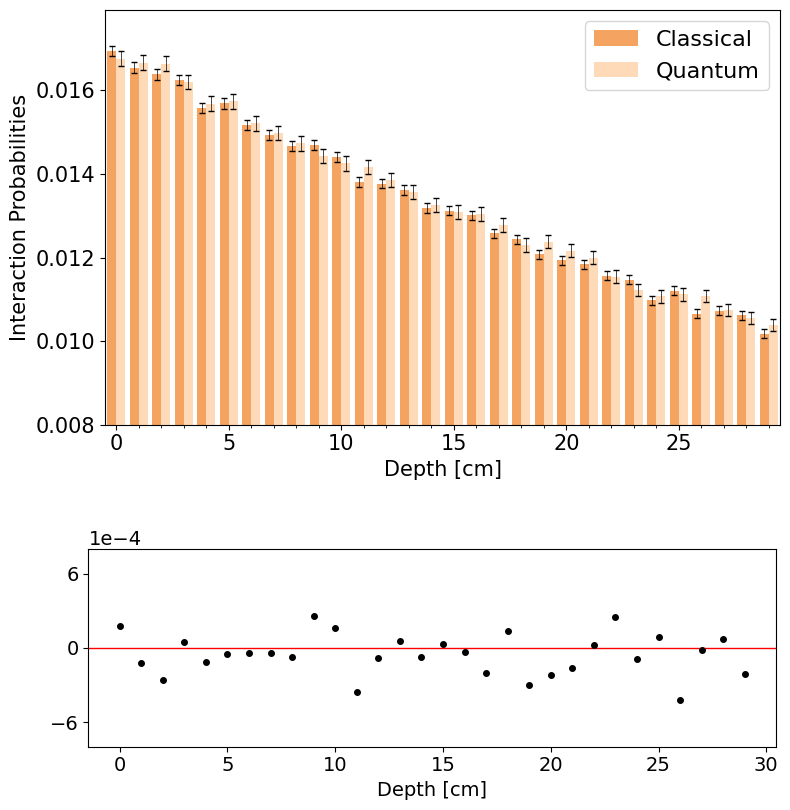}
    \caption{Probability distribution of the 31-step photon interaction simulation for the classical and quantum algorithms. The quantum algorithm has been run on quantum simulator for 500,000 shots, and the classical algorithm
has been run for 1,000,000 shots.}
    \label{fig:two_image}
\end{figure}

\begin{table}[b]
\caption{\label{tab:table1}%
Mean squared error (MSE) and Kullback--Leibler (KL) divergence for 15-step 
and 31-step photon interaction simulations of 10\,MeV photons in a water medium. 
These simulations use the quantum algorithm described in Sec.\,\ref{sec:QW_algorithm} 
and are compared with the classical result.
}
\begin{ruledtabular}
\begin{tabular}{ccc}
\centering
\textrm{Number of step} & \textrm{MSE} & \textrm{KL-divergence} \\
\colrule
15 & \(5.93 \times 10^{-8}\) & \(1.25 \times 10^{-4}\) \\
31 & \(3.03 \times 10^{-8}\) & \(8.20 \times 10^{-5}\) \\
\end{tabular}
\end{ruledtabular}
\end{table}

Fig.~\ref{fig:two_images} and~\ref{fig:two_image} show a comparison of the probability distributions produced by the quantum and classical photon-interaction algorithms with respect to the interaction probabilities. Table~\ref{tab:table1} comparing 15-step and 31-step photon interaction simulations confirms that the QW result is virtually indistinguishable from the classical distribution, at least within the parameter range tested here. Such a match validates the discrete-step quantum walk approach for photon interaction modeling and lays the groundwork for subsequent amplitude estimation. In addition, expanding the simulation steps from 15 to 31 did not result in significant changes in MSE or KL divergence, indicating the scalability of the quantum algorithm. For an $N$-step photon-interaction simulation of a model with $K$ possible interaction channels per branching, a circuit with $K + 2\log_{2}(N+1)$ qubits is required. The total size of the region that can be simulated thus grows exponentially with the number of qubits, whereas the circuit depth increases linearly with the region size. In other words, the proposed quantum-walk-based framework can efficiently simulate realistic photon interactions.

Moreover, to extract meaningful statistical estimates from the above quantum-simulation results, we implement an accelerated approach based on Iterative Quantum Amplitude Estimation (IQAE). IQAE directly extracts the amplitude of the photon survival state $\lvert \psi_1 \rangle$ generated by the quantum random walk, thereby offering a theoretical speed up from $O(1/\epsilon^2)$ to $O(1/\epsilon)$ compared to classical sampling. Due to the memory limitations of the quantum simulator \cite{49_4}, the amplitude estimation was performed on the results of the 15-step photon interaction simulation. In the IQAE simulations, we set the target precision to $\epsilon = 0.01$ and the confidence level to $(1-\alpha)=95\%$, using 30 shots per iteration.

\begin{figure}
    \centering
    \includegraphics[width=0.5\textwidth]{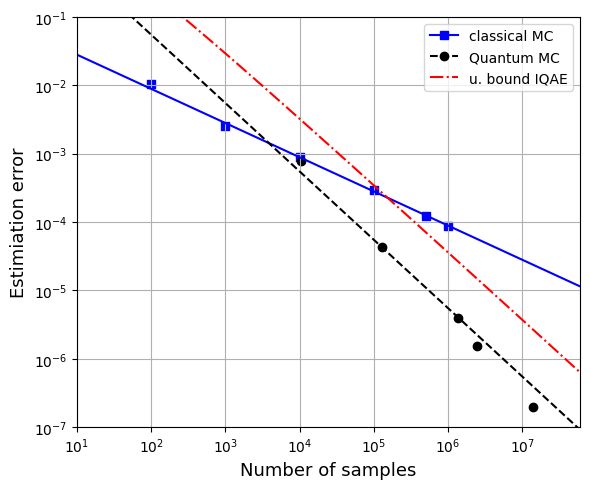}
    \caption{%
    Quantum advantages using the quantum walk based approach with IQAE to estimate probability distribution with quadratic speed up
    }
    \label{fig:result}
\end{figure}

Fig.~\ref{fig:result} shows the estimation accuracy
as a function of the number of oracle queries $N_q$. The estimation accuracy is demonstrated by comparing the absolute error of a quantum algorithm, namely IQAE, against the classical statistical measurement using GEANT4 with respect to the theoretical values. The estimation accuracy \(\epsilon\) of the IQAE exhibits a precise \(O(1/N_q)\) behavior, providing a quadratic speedup compared to the classical method’s \(O(1/\sqrt{N_q})\). This result also fits within the theoretical Chernoff–Hoeffding IQAE bound \cite{40_2}
\begin{equation}
N_{q,\max} 
= \frac{6}{\epsilon}\,
  \log\!\Bigl[
    \frac{2}{\alpha}
    \,\log\!\bigl(\tfrac{\pi}{4\,\epsilon}\bigr)
  \Bigr]
\end{equation}
Such a quadratic speedup can also be observed in other methods like maximum likelihood QAE. Furthermore, through the preceding quantum embedding process, the amplitude of a particular quantum state corresponds to the interaction probability of photons at a specific location. Consequently, the amplitude of that quantum state, as inferred from statistical measurements, carries a direct physical meaning in this context.

%% file: chapter5.tex
\section{Discusssion}
\label{sec:discussion}

In this study, we implemented a simplified model for the interaction of 10 MeV photons in water media using the IBM Qiskit quantum simulator. Compared to classical Monte Carlo methods, our simulation achieves very low mean squared error (MSE) and Kullback–Leibler (KL) divergence values, confirming that the combination of quantum walks and iterative quantum amplitude estimation (IQAE) provides high accuracy and stability in practice. In particular, amplitude estimation via IQAE demonstrates a theoretical quadratic speedup in our experiments.

In comparison to other recent studies, our work contributes in two main aspects. First, most existing research applying QAE often relies on overly simplified models \cite{50_5, 51_5} or quantum generative adversarial network (QGAN)-based methods \cite{52_5, 53_5, 54_5} that do not adequately capture the underlying physical processes. In contrast, our study is the first to introduce a quantum walk (QW) algorithm into the quantum embedding of probability distributions, leading to a more sophisticated and physically meaningful model and experimentally verifying the theoretical quadratic speedup in quantum amplitude estimation. Second, by implementing and scaling up a QW-based interaction model, we confirmed consistent results compared to classical simulations, thereby demonstrating the robust scalability of our method for large-scale high-energy physics (HEP) simulations.

However, the current model employs a simplified approach by assuming only a single interaction channel with complete photon absorption, and thus does not fully reflect complex phenomena such as multiple interactions, partial energy loss, and secondary particle generation in real high-energy physics scenarios. Although this simplification accounts for current hardware constraints, it is important to recognize that realistic high-energy physics requires the inclusion of these more complex interactions. To capture physics beyond the dominant single interaction channel, multiple branching models incorporating additional processes need to be introduced. This can be achieved by expanding the dimension of the coin in the quantum walk, allowing multiple interaction channels within a single step.

Additionally, high-energy photons frequently lose only a fraction of their energy during each interaction and continue to propagate with reduced energy. Consequently, one can discretize the photon energy into multiple levels in the simulation, extending the standard quantum walk formulation to encode additional energy states. However, this approach increases both the number of qubits and the circuit depth. To implement a physical model with probabilistic branches $k$ per step in a quantum circuit, it is necessary to allocate a number of qubits in both the coin register and the walker register that is $\log_2(k)$ times greater than the original requirement. Consequently, incorporating these additional features, given the limitations of current quantum simulator, remains highly challenging.

Incorporating secondary particle generation is also crucial \cite{58_5, 59_5}. One probabilistic branch in the simulation can be designated for secondary particle creation, and an ancilla qubit can be assigned to record the corresponding quantum state. This allows the position, energy, and other parameters of the secondary particle to be tracked. A hybrid quantum-classical scheme enables newly created secondary particles to be simulated independently as new tracks, which can be added to a stack for more realistic simulations.

Since our simulations were conducted under ideal conditions using a quantum simulator, practical constraints such as limited qubit counts, circuit depth, and quantum noise on real quantum hardware were not considered. In particular, our quantum circuit involves multiple multi-qubit gates, which pose challenges due to error accumulation when executed on current quantum hardware \cite{60_5, 61_5}. As a proof-of-concept, we believe that exploring such approaches is essential in anticipation of future advancements in quantum computing. Despite these challenges, our findings demonstrate that the natural integration of Markov-chain-based particle transport processes via quantum walks, combined with advanced amplitude estimation techniques, offers a scalable approach to address large-scale HEP simulation problems.

\section{Conclusion}
\label{sec:conclusion}

In this paper, we propose a new quantum framework that combines discrete-time quantum walk (QW) and iterative quantum amplitude estimation (IQAE) to model high-energy photon transport in water. By using the quantum walk algorithm to embed the discrete stochastic process of photon interactions into quantum superposition states and employing the QAE algorithm to extract the statistical features of this process, we confirm that this approach theoretically guarantees a quadratic speedup over classical Monte Carlo (MC) methods while also enabling highly precise amplitude extraction. Quantum simulation results reveal low mean squared error (MSE) and Kullback–Leibler (KL) divergence, reproducing probability distributions nearly identical to those of classical models. This demonstrates that our framework is both valid and feasible in the field of high-energy physics (HEP).

Although the current model has been simplified to accommodate hardware limitations by adopting a single interaction channel and complete absorption, it provides a robust foundation for incorporating more complex interaction channels and multiple energy levels in the future. As quantum devices advance in terms of qubit counts, coherence time, and gate fidelity, the framework presented in this study is expected to extend to handle increasingly complex physical processes. In addition to photon transport, the same method can be applied to diverse particles and interactions in nuclear and particle physics, offering a powerful tool for accelerating Monte Carlo based analyses in scientific and industrial fields where computational costs are high. Ultimately, this work indicates the significant potential of quantum enhanced algorithms in high-energy physics simulations and opens the door to a scalable solution for overcoming the computational bottlenecks inherent in large-scale MC techniques.